\documentclass[pra,aps,onecolumn,showpacs,amsmath,amssymb]{revtex4}

\usepackage{graphicx}
\usepackage{dcolumn}
\usepackage{bm}
\usepackage{subfigure}
\usepackage{color}
\usepackage{verbatim}

\begin{document}

\title{Evolution of Spin-Orbital Entanglement \\ with Increasing Ising Spin-Orbit Coupling}

\author{Dorota Gotfryd$^{1, 2}$}

\author{Ekaterina P\"arschke$^3$} 

\author{Krzysztof Wohlfeld$^1$} 

\author{Andrzej M. Ole\'s$^{2, 4}$}

\affiliation{$^1$Institute of Theoretical Physics, Faculty of Physics, University of Warsaw, Pasteura 5, PL-02093 Warsaw, Poland}
\affiliation{$^2$Institute of Theoretical Physics, Jagiellonian University, Profesora Stanis\l{}awa {\L}ojasiewicza 11, PL-30348 Krak\'ow, Poland} 
\affiliation{$^3$Institute of Science and Technology Austria, Am Campus 1, A-3400 Klosterneuburg, Austria}
\affiliation{$^4$Max Planck Institute for Solid State Research, Heisenbergstrasse 1, D-70569 Stuttgart, Germany}

\date{\today}

\begin{abstract}
Several realistic spin-orbital models for transition metal
oxides go beyond the classical expectations and could be understood
only by employing the quantum entanglement. Experiments on these
materials confirm that spin-orbital entanglement has measurable
consequences.
Here, we capture the essential features of spin-orbital entanglement
in complex quantum matter utilizing 1D spin-orbital model which
accommodates SU(2)$\otimes$SU(2) symmetric Kugel-Khomskii superexchange
as well as the Ising on-site spin-orbit coupling. Building on the
results obtained  for full and effective models in the regime of strong
spin-orbit coupling, we address the question whether the entanglement
found on superexchange bonds always increases when the Ising spin-orbit
coupling is added. We show that
($i$) quantum entanglement is amplified by strong spin-orbit coupling
and, surprisingly, ($ii$) almost classical disentangled states are
possible. We complete the latter case by analyzing how the entanglement
existing for intermediate values of spin-orbit coupling can disappear
for higher values of this coupling.\\
\textit{Published in:} Condensed Matter {\bf 5}, 53 (2020)
\end{abstract}


\maketitle

\section{Introduction}

Quantum complex matter is characterized by several intertwined degrees
of freedom. Mott insulators with orbital degeneracy belong to this class
of materials and are typically characterized by entangled spin-orbital
states, either only in the excited states or in the ground state as
well. Here we shall present the spin-orbital entanglement in a broader
context, summarizing important results that have been accumulated over
the years. With this background established, we will focus on the
recent study \cite{Got20} of the evolution of spin-orbital entanglement
with increasing Ising spin-orbit coupling, enriching this story
with further details.

By definition, quantum entanglement makes information available in one
part of the quantum state dependable on the result of the measurement
on the other part of the state \cite{Ben06}. A particular kind of
quantum entanglement, involving quantum spin and orbital operators
in the entire quantum system \cite{Ole06} rather than
entanglement between two spatial regions, may appear in the strongly
correlated regime of oxides with orbital degeneracy. The importance of
orbitals, besides spins, in the effective description of such systems
was pointed out long ago by Kugel and Khomskii \cite{Kug82}. After this
pioneering work it has been realized that orbital operators are quantum
in nature, enforcing equal treatment with electron spins and producing
joint spin-orbital fluctuations \cite{Fei97}, both in the ground
\cite{Che07,Brz12} and in excited \cite{You15,Sna19} states.
Quantum fluctuations are further amplified by the spin-orbit coupling
\cite{Jac09}.

In theory, spin-orbital entanglement has been considered in dimerized
phases, spin-orbital models with $S=\frac12$ spins, both in
perovskites \cite{Kha00} and on triangular lattices
\cite{Nor08,Nor11,Cha11}, as well as exactly solvable models such as
SU(2)$\otimes$XY model \cite{Brz14}, but is weakened in layered oxides
suggested as a possible realization of superconductivity \cite{Cha08}.
Several experimental observations in strongly correlated quantum matter
cannot be understood however without spin-orbital entanglement.
For instance, the low energy spectral weight in the optical spectroscopy
for LaVO$_3$ can be explained only when full quantum coupling between
spins and orbitals is included \cite{Miy02,Kha04}, in contrast to
LaMnO$_3$ where orbitals disentangle from large $S=2$ spins
\cite{Kov10}. Entangled states are also found in disordered spin-orbital
liquid states \cite{Lau15,Man18} and in Kitaev quantum liquids
\cite{Bec15,Her18,Rus19,Nat20}, where spatial long-range entanglement
\cite{Tak19} is induced due to strong on-site spin-orbit coupling.

Entanglement in the vanadium perovskites has more manifestations. One
of them are the magnon excitations in the $C$-type AF ($C$-AF) phase of
YVO$_3$, where the gap opens along the $\Gamma$-$Z$ direction
\cite{Ulr03}, indicating the dimerization on the bonds along the $c$
axis which occurs jointly both in spin and orbital channel at finite
temperature \cite{Sir08}. Another experimental confirmation of
spin-orbital entanglement is the evolution of the N\'eel temperature in
the $C$-AF phase with decreasing ionic radius of $R$ ions in $R$VO$_3$
which is induced by the coupling to the orbital state \cite{Hor08}.
Experimentally the entanglement is observed in the critical competition
between the two spin-orbital ordered states \cite{Fuj10} and was
investigated recently in Y$_{0.70}$La$_{0.30}$VO$_3$ single crystals
\cite{Yan19}. Finally, entangled states play a role in the excited
states \cite{CCC} and may help to identify quantum phase transitions in
spin-orbital models \cite{You15}, including the transitions at finite
magnetic field \cite{Tri19}.

Entangled states play also a role in doped spin-orbital systems. One
example is colossal magnetoresistance in manganites, where the orbital
polarons modify the ground state in a one-dimensional (1D) model
\cite{Dag04}. Another example is vanadium perovskites with charged
defects which influence occupied orbitals and lead to the collapse
of orbital order \cite{Ave19}. The spin-orbital entangled states were
also found among the surface states of topological insulators
\cite{Yaj17}. More examples of spin-orbital entanglement could be
found in review articles \cite{Ole12,Brz20}.

As a matter of fact, both superexchange (i.e., intersite) \cite{Ole06}
and spin-orbit coupling (on-site) terms can promote spin-orbital
entanglement, where it may lead to an effective description in terms of
the spin-orbital pseudospins with exotic interactions. The best example
is the onset of the Kitaev-like physics in the iridium oxides
\cite{Jac09} in the so-called relativistic Mott insulators \cite{Wit14}.
Possible competition between these two mechanisms (superexchange bonds
and spin-orbit) or enhancement of entanglement by their joint action is
a challenging problem. It adds another aspect to the complexity of
quantum matter in high-$T_c$ superconductors
\cite{Bia14,Kei15,Cam16,Bus17,Bia18}, intimately connected with quantum
magnetism in strongly correlated transition metal oxides \cite{Kho14}.
Here we analyze the spin-orbital entanglement evolution on a minimal
available model, consisting of SU(2)$\otimes$SU(2) symmetric
superexchange term \cite{Pati} and Ising on-site spin-orbit coupling in
one dimension. Below we recall the recently published thorough numerical
and analytical study \cite{Got20}, supplementing it with the
intermediate global stages and previously not considered cuts, thus
completing the story about the disappearance of spin-orbital
entanglement for certain parameter regime.

\section{The model and basic questions}

\begin{figure}[t!]
\centering
\includegraphics[width=6.5cm]{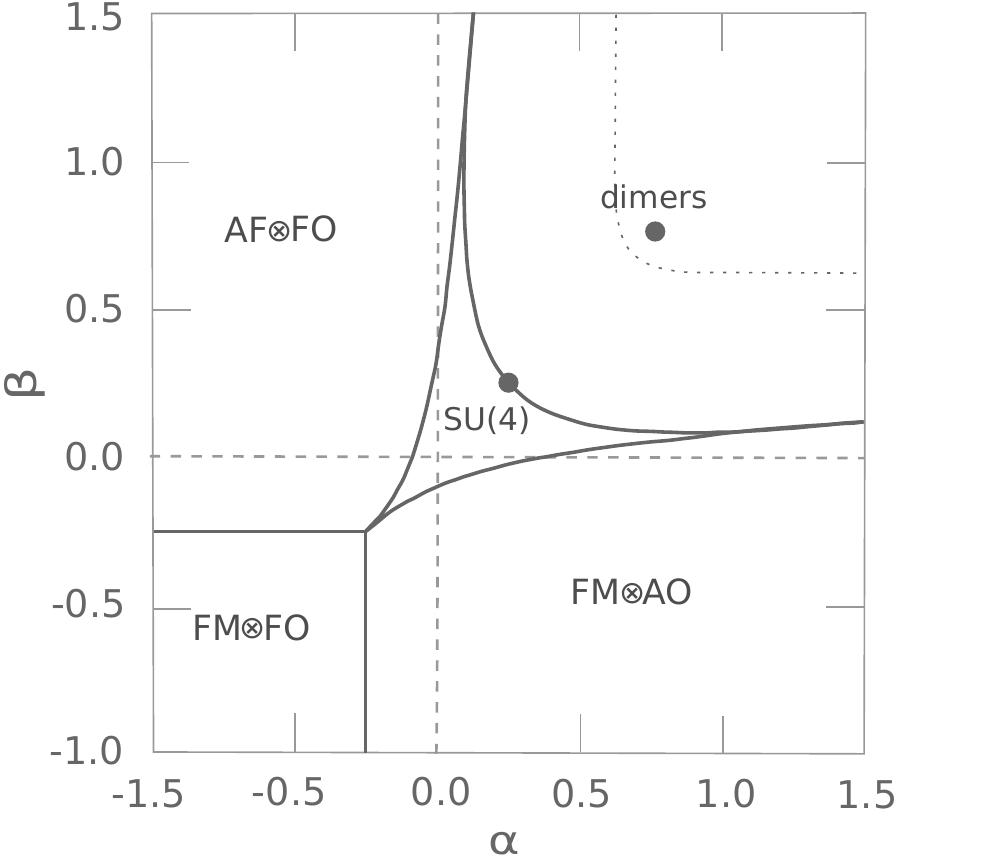}
\caption{Schematic phase diagram of the spin-orbital model (\ref{H_SE})
in the $\{\alpha,\beta\}$ plane inspired by Fig. 1 in \cite{Lun12} and
Fig. 7 in \cite{Got20}.
The model has four distinct phases: product disentangled phases
FM$\otimes$AO, AF$\otimes$FO, FM$\otimes$FO, and the phase around the
SU(4) point with finite spin-orbital entanglement on superexchange bonds.
 }
\label{fig:4}
\end{figure}

The full 1D spin-orbital Hamiltonian considered here has the form
\cite{Got20},
\begin{align}
{\cal H}&={\cal H}_{\text{SE}}+{\cal H}_{\text{SOC}}.
\label{full_H}
\end{align}
The superexchange ${\cal H}_{\text{SE}}$ is a 1D model with two spin
($S=\frac12$) and two orbital (pseudospin) \mbox{($T=\frac12$)}
degrees of freedom per site in a Mott-insulating regime described by
SU(2)$\otimes$SU(2) symmetric expression \cite{Ito00},
\begin{align}
{\cal H}_{\text{SE}}&=J\sum_i\left\{
\left(\textbf{S}_i\cdot\textbf{S}_{i+1}+\alpha\right)
\left(\textbf{T}_i\cdot\textbf{T}_{i+1}+ \beta\right)-\alpha\beta\right\},
\label{H_SE}
\end{align}
where $J\equiv 1$ is the energy unit and the last term $\alpha\beta$
cancels a constant. Spin operators in the scalar product are,
$\{\textbf{S}_i\}\!\equiv\!\{S_i^x,S_i^y,S_i^z\}$,
and similar for pseudospins,
$\{\textbf{T}_i\}\!\equiv\!\{T_i^x,T_i^y,T_i^z\}$. The superexchange
depends on two parameters, $\alpha$ and $\beta$, and the spin/pseudospin
exchange could be of either sign. Hence, also spin [pseudospin] order
could be of either sign, ferromagnetic (FM) or antiferromagnetic (AF)
[ferro-orbital (FO) or alternating-orbital (AO)]. The competition
between different channels of exchange gives a rather rich phase diagram
of Eq. (\ref{H_SE}) \cite{Ito00}, with three product phases,
FM$\otimes$AO, FM$\otimes$FO, and AF$\otimes$FO, see Fig. \ref{fig:4}.
Yet, the entanglement plays a crucial role and the phase diagram
includes the fourth spin-orbitally entangled phase near the exactly
solvable SU(4) point \cite{Ito00}, where the entanglement entropy of the
von Neumann type is finite \cite{Lun12}. Also finite (but much smaller)
spin-orbital entanglement characterizes the parameter range where
$\alpha$ and $\beta$ are large positive, promoting alternating
arrangement of spins (pseudospins) with accompanying quantum
fluctuations~\cite{Lun12}.

The above model (\ref{H_SE}) can be derived from the microscopic
degenerate Hubbard model with the diagonal-only hopping $t$ and the
Coulomb repulsion $U$---then $J=4t^2/U$ and $\alpha=\beta=\frac14$
\cite{Kug15}. The highly symmetric form of spin and pseudospin
interactions follows from the degeneracy of singlet and triplet excited
states. These conditions are not fulfilled in widely studied geometries
with corner-sharing MO$_6$ octahedra (M is a transition metal atom) as
in perovskites, but the model is a realistic Kugel-Khomskii model for
1D chains of face-sharing MO$_6$ octahedra \cite{Kug15}. Such materials
include for example hexagonal crystals such as BaCoO$_3$, BaVS$_3$,
or CsCuCl$_3$. Recently it was shown that the specific heat and the
susceptibility show anomalies at finite temperature which can be
attributed to phase transitions even in the regime of spin-orbital
liquid \cite{Kug19}. Here we treat it as a generic model which stands
for spin-orbital superexchange and entanglement.

As the competing interaction in Eq. (\ref{full_H}) we take Ising
on-site spin-orbit coupling \cite{Got20},
\begin{align}
{\cal H}_{\text{SOC}}&=2\lambda\sum_{i}{S}^z_i{T}^z_i.
\label{soc}
\end{align}
It has $\mathbb{Z}_2$-symmetry and favors the opposite orientation of
spin $\langle{S}^z_i\rangle$ and orbital $\langle{T}^z_i\rangle$
moments at site $i$. This reduced type of spin-orbit coupling occurs,
for instance, when orbital order concerns two $t_{2g}$ orbitals and the
third one is inactive as in LaVO$_3$ or YVO$_3$ \cite{Hor03}. Then, for
an idealized case of untilted oxygen octahedra the spin-orbit coupling
takes the form $2\lambda S^x_iT^y_i$. Surprisingly, similar
$2\lambda S^z_iT^y_i$ form can also be derived for KO$_{2}$, where
peculiar form of superexchange includes $p$ orbitals of molecular oxygen
\cite{Sol08}. Taking the symmetry of Hamiltonian (\ref{full_H}) these
forms are equivalent since they would modify the quantum states in the
same way as the formula in Eq. (\ref{soc}), up to the rotation in the
spin/pseudospin sectors. However, in case of finite orbital-lattice
interactions this should be carefully studied.

For completeness, we add that when the full spin-orbit coupling is
considered \cite{Jac09}, effective entangled pseudospin $\tilde{\bf J}$
operators replace the sum of spin ${\bf S}$ and pseudospin ${\bf T}$---the
$\tilde{\bf J}$ multiplets then define the ground state and excited
states of the system. For instance, spin-orbital entanglement gives the
breakdown of singlets and triplets at the Fermi surface in Sr$_2$RuO$_4$
\cite{Vee14,Bahr,Zha16} and their linear combinations decide about the
orbital character of the pairing components \cite{Ole19}. We believe
that future work will also address the role played by entanglement for
the $d^4$ ions in more detail where the pseudospin could vanish and the
magnetic moments are hidden in the ground state but would become active
in excited states \cite{Kha13,Sou17,Svo17,Li19,Cha19}.

\section{Methods}

\subsection{Entanglement entropy and correlation functions}

In Ref. \cite{Got20} we have verified that for (perturbed) product
phases of the SU(2)$\otimes$SU(2) model (see Fig. \ref{fig:4}), the
spin-orbital entanglement evolution with increasing $\lambda$
experiences only marginal finite-size effects. Here we want to show
the qualitative effects caused by large spin-orbit coupling and for
that the presented studies of small systems suffice. Therefore, we
restrict ourselves to the exact diagonalization (ED) of a 4-site chain
($L=4$) with periodic boundary conditions (PBC). We examine the
spin-orbital entanglement via entanglement entropy of the von Neumann
type per site,
\begin{equation}
\label{vNS}
{\cal S}_{\rm vN}=-\frac{1}{L}\textrm{Tr}_S\{\rho_{S}\ln\rho_{S}\},
\end{equation}
where
\begin{equation}
\rho_{S}=\textrm{Tr}_{T}|\textrm{GS}\rangle\langle\textrm{GS}|
\end{equation}
is the reduced density matrix with the orbital degrees of freedom
traced out, estimating the entanglement by verifying how mixed the
reduced state is. Independently, we quantify spin-orbital entanglement
via correlation on a bond $\langle i,i+1\rangle$ between nearest
neighbors:
\begin{equation}
{\cal C}_{\rm SO}=\frac{1}{L}\sum_{i=1}^{L}
\left\langle(\textbf{S}_{i}\cdot\textbf{S}_{i+1})
(\textbf{T}_{i}\cdot\textbf{T}_{i+1})\right\rangle
-\frac{1}{L}\sum_{i=1}^{L}
\langle\textbf{S}_i\cdot\textbf{S}_{i+1}\rangle\,
\langle\textbf{T}_i\cdot\textbf{T}_{i+1}\rangle.
\label{eq:C}
\end{equation}
At finite $\lambda$, we monitor the response of the system to the Ising
spin-orbit coupling via suitable on-site correlation,
\begin{align}
{\cal O}_{\rm SO}&=\frac{1}{L}\sum_{i=1}^{L}
\left\langle S_i^zT_i^z\right\rangle.
\label{eq:O}
\end{align}
Finally, the spin and orbital states are investigated via spin-only and
orbital-only correlation,
\begin{align}
\label{eq:S}
{\cal S}=&\frac{1}{L}\sum_{i=1}^{L}
\left\langle{\bf S}_{i}\cdot {\bf S}_{i+1}\right\rangle,\\
\label{eq:T}
{\cal T}=&\frac{1}{L}\sum_{i=1}^{L}
\left\langle{\bf T}_{i}\cdot {\bf T}_{i+1}\right\rangle.
\end{align}
They allow to identify the product phases, FM$\otimes$AO, FM$\otimes$FO,
and AF$\otimes$FO in a straightforward way.

The entanglement entropy was investigated previously in the
SU(2)$\otimes$SU(2) model at $\lambda=0$, and was found to be finite
beyond the product phases where at least one (spin or orbital)
component has suppressed quantum fluctuations in a ferro-state
\cite{Lun12}. We have investigated the level of spin-orbital
entanglement in the ground state in the presence of finite spin-orbit
coupling and identified small and large $\lambda$ regimes along the
special $\alpha+\beta=0$ line in the model (\ref{full_H}) \cite{Got20}.
While small spin-orbit terms $\propto\lambda$ can be treated in the
perturbation theory, below we focus on the more interesting regime of
large $\lambda$.

\subsection{Effective Hamiltonian in the large $\lambda$ limit}

In Ref. \cite{Got20}, following the general scheme set in Ref.
\cite{Jac09}, we have described the low-energy physics in the
$\lambda\rightarrow\infty$ limit by the model utilizing effective
pseudospins $\{\tilde{\textbf{J}}_i\}$. Namely, the Hamiltonian
(\ref{H_SE}) projected onto the low-energy part of
${\cal H}_{\text{SOC}}$ eigenbasis takes the XXZ form,
\begin{align}
{\cal H}_{\rm eff}=\frac12\, J \sum_{i} \left\{\,
 \tilde{\text{J}}^{x}_{i}\,\tilde{\text{J}}^x_{i+1}
+\tilde{\text{J}}^{y}_{i}\,\tilde{\text{J}}^y_{i+1}
+2(\alpha+\beta)\,\tilde{\text{J}}^z_i\,\tilde{\text{J}}^z_{i+1}\,\right\}.
\label{Heff}
\end{align}
It is remarkable that the Ising part of the above Hamiltonian,
$2(\alpha+\beta)\,\tilde{\text{J}}^z_i\,\tilde{\text{J}}^z_{i+1}$,
vanishes and changes sign at $\alpha+\beta=0$. As a result one finds
the XY model at $\beta+\alpha=0$, AF Heisenberg model at
$\alpha+\beta=\frac12$, and hidden FM Heisenberg at the line
$\alpha+\beta=-\frac12$,
denoted by the white dashed lines on Fig. \ref{fig:1}(\textbf{c}).
Below the latter line one finds a FM Ising ground state, with
$\langle\tilde{\text J}^z_i\,\tilde{\text J}^z_{i+1}\rangle=\frac14$
and vanishing quantum fluctuations,
$\langle\tilde{\text{J}}^x_i\,\tilde{\text{J}}^x_{i+1}\rangle=
 \langle\tilde{\text{J}}^y_i\,\tilde{\text{J}}^y_{i+1}\rangle=0$. For
this state one finds that the mean field approximation for the spin-spin
correlations is exact and ${\cal C}_{\text SO}=0$ (\ref{eq:C}), i.e.,
spin-orbital entanglement vanishes.

In Fig. \ref{fig:1}(\textbf{c}) the highly entangled state exists in a
close neighborhood of the $\beta+\alpha=0$ line. More precisely, highly
entangled 1D spin-liquid class of states, called also 1D quantum
antiferromagnet in the literature, could be found in a "quantum stripe"
with one sharp boundary at $\alpha+\beta=-\frac12$ and system
size-dependent soft boundary seemingly approaching AF Heisenberg line
$\alpha+\beta=\frac12$ for larger chains, thus roughly fitting
within the two outer white dashed lines. We observe that the
entanglement is maximal between the $\alpha+\beta=-\frac12$ and
$\alpha+\beta=0$ lines, and decays at a rate depending on the system
size. This behavior is verified for finite systems, i.e., for rings of
$L=4$ sites (shown here) and of $L=8,12$ sites (studied before in
\cite{Got20}). From $\alpha+\beta=\frac12$ upwards we find
interpolation between the 1D Heisenberg antiferromagnet and the
classical N\'eel state in the limit of $\alpha+\beta\rightarrow\infty$
where the quantum fluctuations are quenched.

\section{Results and discussion}

\subsection{Gradual evolution of spin-orbital entanglement
            with increasing Ising spin-orbit coupling}

\begin{figure}[t!]
\centering
\includegraphics[width=16.0cm]{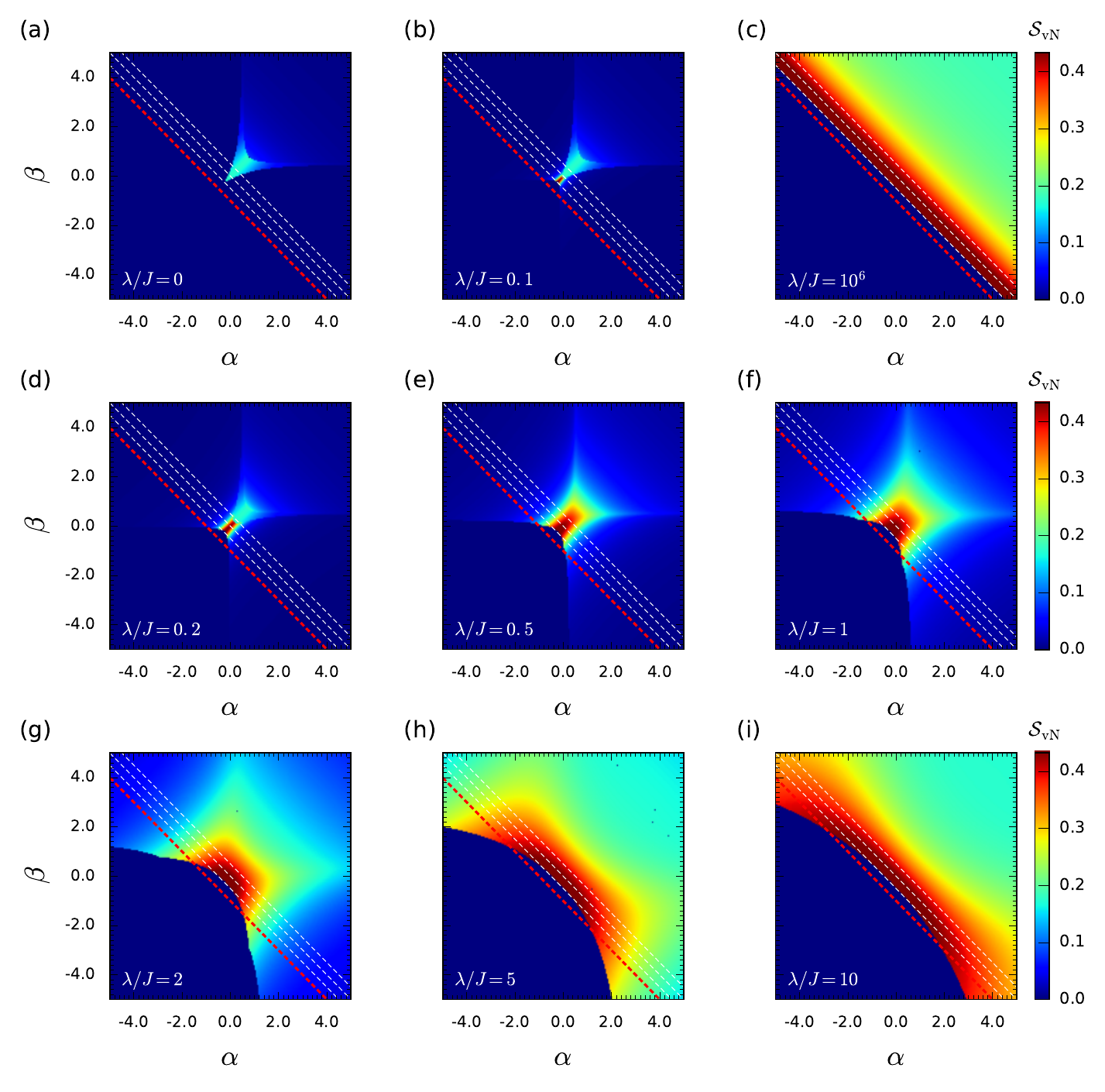}
\caption{Evolution of the entanglement entropy ${\cal S}_{\rm vN}$
(\ref{vNS}) with increasing $\lambda$ for the Hamiltonian
(\ref{full_H}) calculated for a ring of $L=4$ sites. Panels present
slices of the $\{\alpha,\beta\}$ parameter space for a fixed value of
$\lambda/J$. Panels (\textbf{a})--(\textbf{c}) reproduce surprisingly
well the result presented in Fig. 1 of Ref. \cite{Got20} for $L=12$.
Panels (\textbf{d})--(\textbf{i}) reveal intermediate stages between
the panels (\textbf{b}) and (\textbf{c}). The three white dashed lines
represent (from left to right) hidden FM Heisenberg, XY, and AF
Heisenberg models emerging in the $\lambda\rightarrow\infty$ limit at
$\alpha+\beta=-\frac12$, $\alpha+\beta=0$, and
$\alpha+\beta=\frac12$ (white dashed lines from bottom to top).
For longer chains, e.g. $L=12$, the "quantum stripe" (red region near
the $\alpha+\beta=0$ line) in panel (\textbf{c}) fits between hidden
FM Heisenberg and AF Heisenberg lines. Red dashed lines in the panels
indicate the cross section along which the $\{\alpha,\lambda\}$ maps
are calculated in Figs. \ref{fig:2} and \ref{fig:3}.
 }
\label{fig:1}
\end{figure}

Here we first complement the analysis of Ref. \cite{Got20} with the
data for the entanglement entropy in the intermediate regime of
$\lambda/J$, building a global picture of spin-orbital entanglement
under increasing $\lambda$. Namely, we add 6 entanglement entropy scans
in Figs. \ref{fig:1}(\textbf{d})--\ref{fig:1}(\textbf{i}), fitting in
between the panels (\textbf{b}) and (\textbf{c}). Results presented
here for $L=4$ include small finite size effects and thus one finds
finite spin-orbital entanglement above the stripe for large $\lambda$.
Nevertheless, the evolution of perturbed product phases is
qualitatively the same as before for all $4n$ chains \cite{Got20},
showing that spin-orbital entanglement is pretty local.

Finite spin-orbital entanglement in the central region of $\lambda=0$
slice [Fig. \ref{fig:1}(\textbf{a})] increases and spreads out with
growing $\lambda$, see panels (\textbf{b}) and
(\textbf{d})--(\textbf{g}). In particular, infinitesimal values of
entanglement entropy appear on the boundaries of perturbed product
phases containing either AF or AO component, see
panels (\textbf{b}), (\textbf{d}), and clearly visible on panels
(\textbf{e})--(\textbf{f})). At the same time, FM$\otimes$FO phase does
not show any entanglement, which cannot be induced in this classical
state by also classical spin-orbit Ising coupling (\ref{soc}).
Interestingly, the border between completely classical region,
originally hosting FM$\otimes$FO ground state, and the spin-orbitally
entangled region gradually changes its shape. The change starts for
infinitesimal $\lambda>0$ but can be resolved starting from panel
(\textbf{d}). A sharp boundary, at first restricting the classical state
to the third quarter of panel (\textbf{b}), moves so that a much wider
classical region is visible in panel (\textbf{i}). This result is quite
close to the classical triangle seen in panel (\textbf{c}).

There is a huge difference in size between the maximal and minimal
spin-orbital entanglement for varying $\lambda$, displayed in Figs.
\ref{fig:1}(\textbf{a})--\ref{fig:1}(\textbf{b}), and
\ref{fig:1}(\textbf{c}), correspondingly.
At the $\lambda=0$ case the spin-orbital term,
$(\text{\textbf{S}}_i\text{\textbf{S}}_{i+1})
 (\text{\textbf{T}}_i\text{\textbf{T}}_{i+1})$,
can produce substantial entanglement in the ground state only in a
small approximately triangular region, where $\beta$ and $\alpha$
scales of purely spin or purely orbital interactions are small enough.
This corresponds to the critical spin-orbitally entangled region in the
phase diagram of the SU(2)$\otimes$SU(2) model \cite{Ito00}.
In $\lambda\rightarrow\infty$ limit, the "quantum stripe" characterized
by the highly entangled state can be also related to the
$(\text{\textbf{S}}_i\text{\textbf{S}}_{i+1})
 (\text{\textbf{T}}_i\text{\textbf{T}}_{i+1})$
joint interaction. Namely, when projecting the Hamiltonian (\ref{H_SE})
on the lower part of ${\cal H}_{\text{SOC}}$ eigenbasis term by term,
one observes that
$(\text{\textbf{S}}_i\text{\textbf{S}}_{i+1})
 (\text{\textbf{T}}_i\text{\textbf{T}}_{i+1})$
reduces to the XY term, independent of $\alpha$ and $\beta$, while
 $\beta\,\text{\textbf{S}}_i\text{\textbf{S}}_{i+1}$ and
$\alpha\,\text{\textbf{T}}_i\text{\textbf{T}}_{i+1}$ transform into
 $\propto\beta\,\tilde{\text{J}}^{z}_i\tilde{\text{J}}^{z}_{i+1}$ and
$\propto\alpha\,\tilde{\text{J}}^{z}_i\tilde{\text{J}}^{z}_{i+1}$,
respectively. As a result, on the $\alpha+\beta=0$ line the Ising
terms cancel out, leaving the XY model with entangled ground state,
and on the entire "quantum stripe" the ratio Ising/XY supports the
highly entangled state. The left boundary of the stripe is well defined
by the $\alpha+\beta=-\frac12$ line where FM Heisenberg ground
state appears. The degenerate energy splits immediately below the
above line, leaving Ising 2-fold degeneracy.

\subsection{Entanglement evolution for the quantum phases along
            the $\alpha+\beta=0$ line}

\begin{figure}[t!]
\centering
\includegraphics[width=16cm]{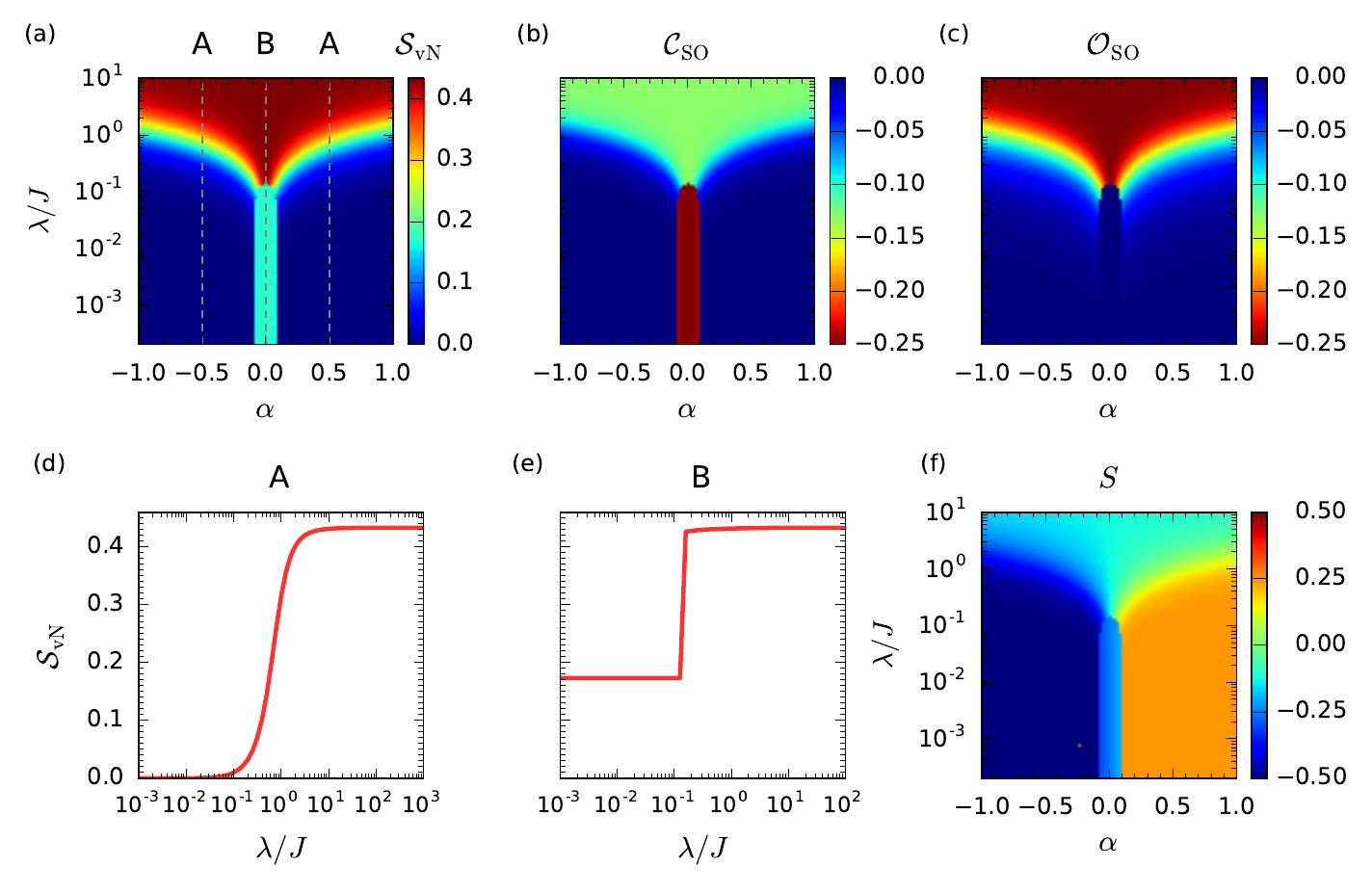}
\caption{
Evolution of the von Neumann spin-orbital entanglement entropy and the
three spin-orbital correlation functions in the ground state of
Hamiltonian (\ref{full_H}) with $\alpha+\beta=0$ for
$\alpha\in[-1.0,1.0]$, calculated with ED for the ring of $L=4$ sites
for logarithmically increasing spin-orbit coupling $\lambda$:
(\textbf{a})---the von Neumann spin-orbital entanglement entropy
    ${\cal S}_{\rm vN}$ (\ref{vNS});
(\textbf{b})---the intersite spin-orbital correlation function
    ${\cal C}_{\rm SO}$ (\ref{eq:C});
(\textbf{c})---the on-site spin-orbit correlation function
    ${\cal O}_{\rm SO}$ (\ref{eq:O});
(\textbf{f})---the spin correlation function ${\cal S}$ (\ref{eq:S}).
Panels (\textbf{d}) \& (\textbf{e}) show the von Neumann spin-orbital
entanglement entropy ${\cal S}_{\rm vN}$ (\ref{vNS}) obtained for
increasing $\lambda$ at $|\alpha|=\frac12$ [cut A in ((\textbf{a})]
and fitted with a logistic function (black thin line), and at
$\alpha=0$ [cut B in (\textbf{a})], respectively.}
\label{fig:2}
\end{figure}

Next, we recall the evolution of spin-orbital entanglement on the
$\alpha+\beta=0$ line, marked by the middle diagonal white dashed line
in different panels of Fig. \ref{fig:1} and investigated originally in
Ref. \cite{Got20} (see Fig. 2 there).
This line goes through the two product phases,
FM$\otimes$AO and AF$\otimes$FO, and through the entangled phase which
separates them (similar to the white line in Fig. \ref{fig:1} at
$\alpha+\beta=\frac12$). Two distinct types of evolution emerged
on the vertical lines in Fig. \ref{fig:2} (under increasing $\lambda$):
($i$) evolution of type A starting from AF$\otimes$FO or FM$\otimes$AO
state with no spin-orbital entanglement at $\lambda=0$, and
($ii$) evolution of type B starting from the spin-orbitally entangled
critical SU(4) singlet ground state \cite{Lun12}. Altogether, the
behavior of the entanglement entropy and the correlation functions
displayed in Fig. \ref{fig:2} for the $L=4$ system is very similar to
that shown in Fig. 2 of Ref. \cite{Got20} for the $L=12$ system which
confirms that especially for the evolution of type A the spin-orbital
entanglement is a local phenomenon and the finite size effects are
small in the quantities analyzed here.

In case ($i$), the ground state has large degeneracy at $\lambda=0$ due
to FM or FO sector. From the energetic point of view the level crossing
is at $\lambda=0$ in the degenerate state. Then $\lambda\neq0$ lifts the
degeneracy and leads to non-degenerate perturbed product states. From
that on, all changes in the ground state have a crossover character
which is visible in the smooth increase of the entanglement entropy
${\cal S}_{\rm vN}$ along the line A, see Fig. \ref{fig:2}(\textbf{d}).
Finite entanglement is then transferred to the product phases at
increasing $\lambda$. But the most important is that even when the
energy scales fully separate at $\lambda/J\gg 1$, finite spin-orbital
entanglement is found on superexchange bonds, see Fig.
\ref{fig:2}(\textbf{b}). The finite value of
${\cal C}_{\text{SO}}$ (\ref{eq:O}) is induced by the
$\left(\textbf{S}_i\cdot\textbf{S}_{i+1}+\alpha\right)
 \left(\textbf{T}_i\cdot\textbf{T}_{i+1}+ \beta\right)$ term to be
reduced to the XY model and freed from $\alpha$ and $\beta$-dependent
Ising contributions in $\lambda\rightarrow\infty$ limit. Figure
\ref{fig:2} shows that this limit is approximately valid as soon as
the non-zero entropy and ${\cal C}_{\text{SO}}$ values spread on
$\alpha+\beta=0$ line. This can be also verified by the maximal
response in on-site correlations ${\cal O}_{\text{SO}}$ (\ref{eq:O}),
see Fig. \ref{fig:2}(\textbf{c}).

\begin{figure}[t!]
\centering
\includegraphics[width=16cm]{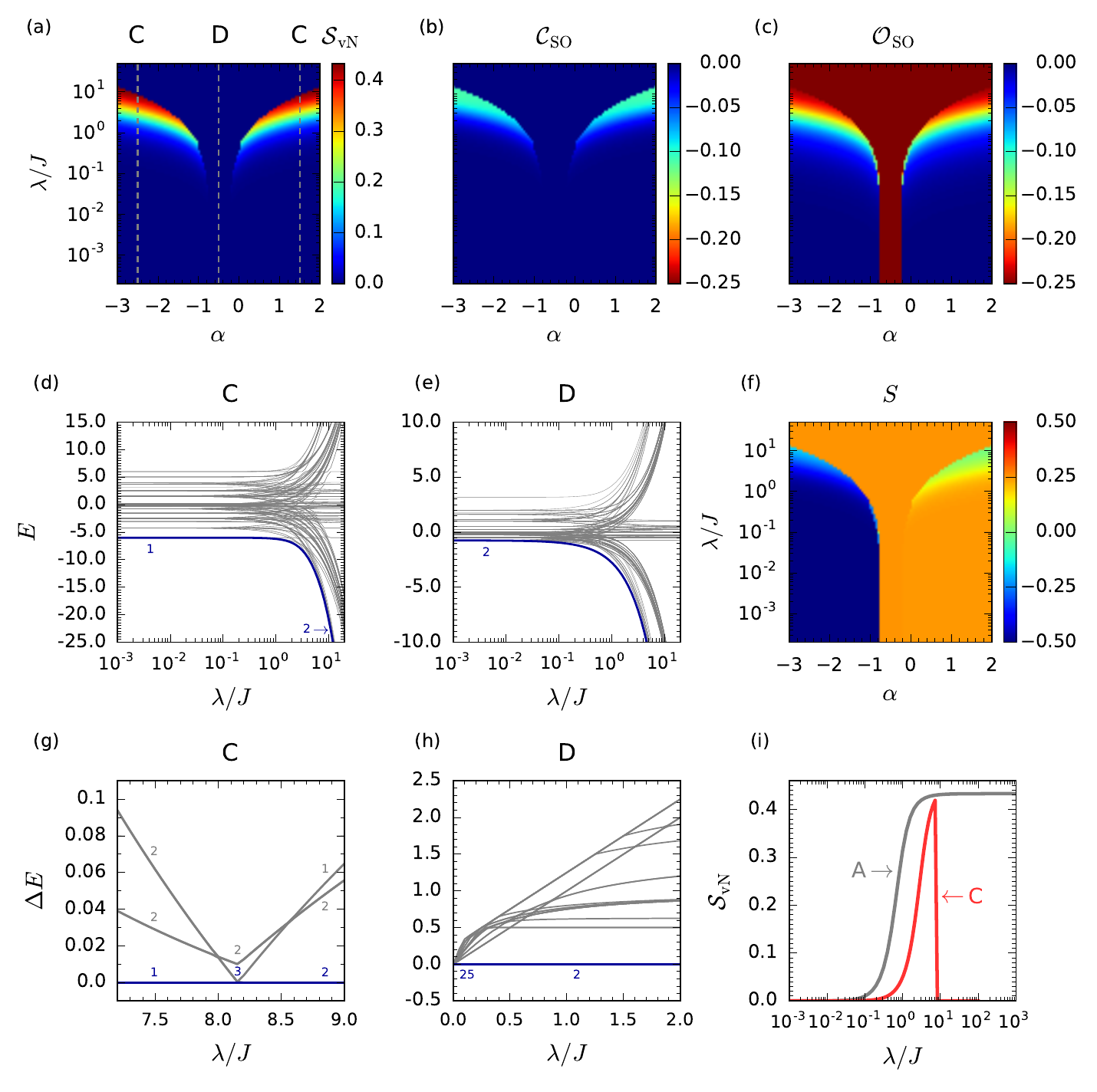}
\caption{Absence of large spin-orbital entanglement along the red
$\alpha+\beta=-1$ line in Fig. \ref{fig:1} for the 1D ring of $L=4$
sites.
Top panels show the maps for the $(\alpha,\lambda)$-plane:
(\textbf{a})---entanglement entropy ${\cal S}_{\text{vN}}$ (\ref{vNS});
(\textbf{b})---spin-orbital ${\cal C}_{\rm SO}$ correlation (\ref{eq:C}),
(\textbf{c})---on-site ${\cal O}_{\rm SO}$ correlation (\ref{eq:O}), and
(\textbf{f})---spin-only correlation ${\cal S}$ (\ref{eq:S}).
Panels (\textbf{d}) \& (\textbf{e}) present full energy spectrum along
cross section C (D), with the ground state energy marked in blue.
Panel (\textbf{g})---the energy differences $\Delta E=E_i-E_0$ between
the first four excited states and the ground state around the sharp
transition from non-degenerate perturbed AF$\otimes$FO (FM$\otimes$AO)
level into 2-fold degenerate Ising FM ground state.
Panel (\textbf{h}) presents the splitting of 25-fold degenerate
FM$\otimes$FO ground state at infinitesimal $\lambda>0$---it gives the
2-fold degenerate Ising FM ground state.
In panel~(\textbf{i}) we contrast the entanglement enetropy
${\cal S}_{\text{vN}}$ (\ref{vNS}) along the line C and the line A
in Fig. \ref{fig:2}.}
\label{fig:3}
\end{figure}

Entanglement entropy ${\cal S}_{\rm vN}$ in case ($ii$) starts from a
finite value at $\lambda\to 0$ in spin-orbitally entangled critical
singlet phase \cite{Lun12}. Such singlet correlations are robust along
the path B and thus the spin-orbital entanglement (described by the
entropy and the nearest neighbor correlations ${\cal C}_{\text{SO}}$)
can change only in a discontinuous way, by the kink shown in Fig.
\ref{fig:2}(\textbf{e}). We have seen before that the evolution of
spin-orbital entanglement for $4n$--site chains occurs along the line
of type B through a series of $n$ kinks \cite{Got20}. Entanglement
gradually gets more robust up to $\lambda_{crit}/J\simeq 0.2$, above
which the highly-entangled state is reached. With this, we completed
a short summary of the main results of Ref. \cite{Got20}.

\subsection{Entanglement evolution for the quantum phases along
            the $\alpha+\beta=-1$ line}

Figure \ref{fig:1} revealed the gradual evolution of the spin-orbital
entanglement in $\lambda$. Moreover, we have seen how  classical region
in both spin and orbital sectors, initially restricted to FM$\otimes$AO
phase in the third quarter of the diagram, gradually spreads out,
reaching extended triangular shape within the boundaries shown in Fig.
\ref{fig:1}(\textbf{c}). Within this triangular region, the evolution
of FM$\otimes$FO phase with increasing $\lambda$ is rather trivial as
long as we have the Ising form of the spin-orbit coupling (\ref{soc}).
However, it is not obvious how the lower parts of AF$\otimes$FO or
FM$\otimes$AO phases, see Fig. \ref{fig:4}, approach the Ising FM
ground state below the $\alpha+\beta=-\frac12$ line
(the lowest white dashed line in Fig. \ref{fig:1}).
Here, we concentrate on the evolution in $\lambda$ along the
representative $\alpha+\beta=-1$ line (red dashed line in Fig.
\ref{fig:1}), including both of the above cases, completing the review
of the $\lambda$ evolution for the Hamiltonian Eq. (\ref{full_H}).

We have selected this line because it crosses all three product phases
AF$\otimes$FO, FM$\otimes$FO, and FM$\otimes$AO at $\lambda=0$ (Fig.
\ref{fig:4}) and lies below the "quantum stripe" for large $\lambda$.
The line runs close enough to the center of the $\{\alpha,\beta\}$ maps
to bear quite insightful information. The entanglement evolution in
$\lambda$ along the line is shown in Fig. \ref{fig:3}. Panels
(\textbf{a}) and (\textbf{b}) show, respectively, the entanglement
entropy and the ${\cal C}_{\text{SO}}$ maps. One can observe that
AF$\otimes$FO and FM$\otimes$AO, represented again by cross section C,
at first gain gradually spin-orbital entanglement with growing
$\lambda$, just like in the evolution along the line A in Fig.
\ref{fig:3}. ${\cal S}$ and ${\cal T}$ correlations confirm that those
are indeed perturbed AF$\otimes$FO and FM$\otimes$AO phases. [${\cal S}$
correlation is shown in Fig. \ref{fig:3}(\textbf{f}).] However, for some
particular value of $\lambda$ the growth of entanglement is sharply cut
off. For parallel lines lying deeper within FM$\otimes$FO phase, the
effects discussed here will appear as well but for higher values of
$\lambda$ (see the trends in Fig. \ref{fig:1}).

The origin of this evolution can be resolved by inspecting the energy
spectrum of the $L=4$-site ring, plotted in Fig. \ref{fig:3}(\textbf{d})
at $\alpha=-2.5$ and $\beta=1.5$ and varying $\lambda$. Here we note
that for $\lambda>0$ the ground state degeneracy is completely lifted.
It is quite difficult to spot the level crossing suggested by sharp
changes in entanglement. Therefore, in Fig. \ref{fig:3}(\textbf{g}) we
analyzed the energy differences between the excited energies and the
ground state energy, $\Delta E=E_{i}-E_{0}$, plotted in gray. For
convenience the ground state energy is renormalized as
$\Delta E=E_i-E_0=0$ and plotted in blue. One observes that at
$\lambda\approx8.175\pm 0.025$, 2-fold degenerate excited level crosses
with the non-degenerate ground state, leaving 2-fold degenerate ground
state. In this way the perturbed product state becomes an excited state
and gradually gains entanglement, even crossing another excited 2-fold
degenerate state for further increasing $\lambda$, while the Ising
ferromagnet becomes the ground state.

For completeness, we describe also the evolution of the FM$\otimes$FO
state, choosing cross section B at $\alpha=\beta=-\frac12$, see Fig.
\ref{fig:3}(\textbf{h}). The degenerate 25-fold FM$\otimes$FO
level splits immediately when $\lambda>0$, leaving only 2-fold
degeneracy. Consistently, panels (\textbf{a}) and (\textbf{b}) confirm
no spin-orbital entanglement, while panels (\textbf{c}) and (\textbf{h})
show that the response of the system to spin-orbit coupling is
instantaneous.

\section{Summary and conclusions}

In summary, both in Ref. \cite{Got20} and here, we have accumulated
evidence for two distinct regimes of Ising spin-orbit coupling:
($i$)~when it is weak, i.e., $\lambda<\lambda_{crit}$, one is in the
perturbative regime and the superexchange is almost undisturbed, while
($ii$) strong spin-orbit coupling $\lambda>\lambda_{crit}$ reduces
entanglement due to superexchange but may induce it in the product
phases where quantum fluctuations are finite at $\lambda=0$, in either
spin or orbital channel. The fluctuations enhance strongly the
entanglement of already spin-orbital entangled central region. Note
that here we define $\lambda_{crit}$ differently for the evolution of
type A and B in Fig. \ref{fig:2} and type C and D in Fig. \ref{fig:3}.

For the evolution of type A on the $\alpha+\beta=0$ line,
$\lambda_{crit}$ stands for the dynamical region of rapid entanglement
growth (in logarithmic scale), see Fig. \ref{fig:2}. It connects the
perturbed FM$\otimes$AO and AF$\otimes$FO phases to the highly
entangled state at $\lambda>\lambda_{crit}$ via crossover \cite{Got20}.
The evolution of type C on the $\alpha+\beta=-1$ line is more subtle.
The perturbed FM$\otimes$AO (AF$\otimes$FO) rapidly gains entanglement
in a smooth way but the crossover process is interrupted by a level
crossing with the Ising ferromagnet. These transitions seem to be
immune to finite-size effects.

For the evolution of type B on the $\alpha+\beta=0$ line, analyzing the
spin-orbital entangled critical SU(4) singlet state through $n$ kinks
for $L=4n$-site system, we can set
$\lambda_{crit}(L\rightarrow\infty)\approx0.2 J$ \cite{Got20}. In other
words, finite-size scaling analysis in Ref. \cite{Got20} suggests that
the entanglement growth from the singlet state will be continuous in
the thermodynamic limit, with a phase transition at
$\lambda_{crit}\approx0.2J$. Formally $\lambda_{crit}$ could be set to
0, since the large FM (FO) ground state degeneracy at $\lambda=0$ is
immediately reduced to 2-fold degeneracy at $\lambda=0$, matching the
Ising state. However, the simplest possible state representing the
ground state energy is disentangled in the whole $\lambda$ range.
The case of FM$\otimes$FO is rather special as the quantum fluctuations
are suppressed in both spin and orbital channel at $\lambda=0$. They
are also absent in the Ising spin-orbit coupling (\ref{soc}), so
irrespectively of the actual value of $\lambda$, there is no driving
force to generate entangled states. As a result, this phase remains
disentangled and has just maximal on-site spin-orbital correlations
${\cal O}_{\text SO}=-\frac14$ which are unable to induce any
intersite entanglement.

We emphasize that this will change when spin-orbit coupling is quantum
and many-body states fully develop. For instance, such states are
found in two-dimensional (2D) iridates Ba$_2$IrO$4$ and Sr$_2$IrO$_4$,
where it was demonstrated that electron- and hole-doped iridates are
fundamentally different \cite{Par17}. A similar situation is found in
CuAl$_2$O$_4$ spinel, where spin-orbital entangled Kramers doublets
form and again the electron addition spectra are well described by the
$J_{eff}=\frac12$ hole state, whereas electron-removal spectra have
a rich multiplet structure \cite{Kim19}.
Although the low-energy physics depends on the iridate geometry, it
would be challenging to investigate in a similar way the interplay of
spin-orbital superexchange with quantum spin-orbit coupling.

We also reported an interesting "emergence" of the spin-orbital
entanglement along the $\alpha+\beta=-1$ line in the Ising limit. In
this case, both for the negligibly small $\lambda$ as well as for the
large $\lambda>\lambda_{crit}$ the spin-orbital entanglement vanishes.
Surprisingly, however, for a moderate value of $\lambda$ (and a
finite value of $\alpha$) the spin-orbital entanglement becomes
finite. This shows the intricate complexity of the studied system.
Crucially, the analysis at $\lambda>\lambda_{crit}$ is consistent with
the phase diagram of the effective XXZ Hamiltonian (\ref{Heff}). In
particular, the ground state with vanishing spin-orbital entanglement
appears when the quantum fluctuations vanish in the ground state of an
effective model. This is the case of an Ising ferromagnet obtained at
$\alpha+\beta\le -\frac12$.

Finally, we argue that the results presented here may play an important
role (after extension) in the understanding of the strongly correlated
systems with non-negligible spin-orbit coupling---such as, e.g. the $5d$
iridates, $4d$ ruthenates, $3d$ vanadates, the $2p$ alkali hyperoxides,
and other to-be-synthesized materials. More discussion of the
entanglement in these materials was given before \cite{Got20}.
As discussed also in Ref. \cite{Got20}, the results presented for the 1D
model can be (on a qualitative level) extended to the higher-dimensional
hypercubic lattice. This is due to the fact that the mapping to the
effective model is valid independently of the dimension. Therefore, the
spin-orbital correlation function (and consequently the spin-orbital
entanglement) is nonzero for the similar range of model parameters in
2D or three-dimensional systems as for the 1D model reported here.

\vspace{6pt}

{\it Author contributions:}
All authors participated equally in the formulation of research ideas;
calculations, D.G.;
derivation of the effective strong coupling model, E.M.P.;
results selection and paper writing, D.G. and A.M.O.;
graphical visualization, D.G.;
ultimate formulation--review, citations, and correspondence, A.M.O.

{\it Funding:} This research was supported by Narodowe Centrum Nauki
(NCN, Poland) under Projects
No. 2016/23/B/ST3/00839 and No. 2016/22/E/ST3/00560.
E.M.P. was supported by the European Union's Horizon 2020
research and innovation programme under the Maria
Sk\l{}odowska-Curie grant agreement No. 754411.

{\it Acknowledgments:}
It is our great pleasure to thank Wojtek Brzezicki, Ji\v{r}\'i
Chaloupka, Daniel I. Khomskii, and Klim Kugel for many insightful
discussions. A.~M.~Ole\'s is grateful for the Alexander von Humboldt
Foundation Fellowship \mbox{(Humboldt-Forschungspreis).}

\end{document}